\documentclass[%
 aip,
 apl,
 sd,%
amsmath,amssymb,
 reprint%
]{revtex4-1}

\usepackage{graphicx}
\usepackage{dcolumn}
\usepackage{bm}

\begin{document}

\preprint{AIP/123-QED}

\title[]{Measurement of Ultrafast Dynamics of Photoexcited Carriers in $\beta$-Ga$_2$O$_3$ by Two-Color Optical Pump-Probe Spectroscopy}

\author{Okan Koksal}
 \email{ok74@cornell.edu}

\author{Nicholas Tanen}%

\affiliation{ 
	School of Electrical and Computer Engineering, Cornell University, Ithaca, NY 14853, USA. 
}%

\author{Debdeep Jena}
\affiliation{ 
	School of Electrical and Computer Engineering, Cornell University, Ithaca, NY 14853, USA. 
}%
\affiliation{%
	Department of Materials Science and Engineering, Cornell University, Ithaca, NY 14853, USA.
}%

\author{Huili (Grace) Xing}
\affiliation{ 
	School of Electrical and Computer Engineering, Cornell University, Ithaca, NY 14853, USA. 
}%
\affiliation{%
	Department of Materials Science and Engineering, Cornell University, Ithaca, NY 14853, USA.
}%

\author{Farhan Rana}

\affiliation{ 
	School of Electrical and Computer Engineering, Cornell University, Ithaca, NY 14853, USA. 
}%

\date{\today}

\begin{abstract}
We report results from ultrafast two-color optical pump-probe spectroscopy on bulk $\beta$-Ga$_2$O$_3$. A two-photon absorption scheme is used to photoexcite carriers with the pump pulse and free-carrier absorption of the probe pulse is used to record the subsequent dynamics of the photoexcited carriers. Our results are consistent with carrier recombination via defect-assisted processes. We also observe transient polarization-selective optical absorption of the probe pulse by defect states under nonequilibrium conditions. A rate equation model for electron and hole capture by defects is proposed and used to explain the data. Whereas the rate constants for electron capture by defects are found to be temperature-independent, they are measured to be strongly temperature-dependent for hole capture and point to a lattice deformation/relaxation process accompanying hole capture. Our results shed light on the mechanisms and rates associated with carrier capture by defects in $\beta$-Ga$_2$O$_3$.  
\end{abstract}

\maketitle

Ga$_2$O$_3$ polymorphs, most notably $\beta$-Ga$_2$O$_3$, have been the object of renewed interest in recent years. The wide bandgap ($\sim$4.7 eV) of $\beta$-Ga$_2$O$_3$ holds tremendous promise for high power electron devices~\cite{ii_higa16,mast17,pear18,xing18} and solar-blind photon detectors~\cite{hu15,soo16}. $\beta$-Ga$_2$O$_3$ has a large number of intrinsic and extrinsic defects and trap states with unusual and poorly understood properties that have a direct bearing on the performance of devices. These include intrinsic defects due to gallium and oxygen vacancies~\cite{Frau2017,Ingeb2018,Zhang2016,Dong2017,Varley2010,Martin2013}, impurity atoms that act as shallow and deep donors and deep acceptors~\cite{Ingeb2018,Zhang2016,Peel2016,Neel2018}, and crystal dislocations and nanopipes~\cite{Nakai2015}. Defect states in $\beta$-Ga$_2$O$_3$ can act as very efficient centers of carrier capture and emission and can contribute to both radiative and non-radiative carrier recombination. Understanding the mechanisms of carrier capture by these defects and the associated rates and time scales are critical for the development of $\beta$-Ga$_2$O$_3$ technologies. Several experimental investigations of shallow and deep defects have been reported in the literature using different techniques such as DLTS/DLOS~\cite{Zhang2016}, photoluminescence (PL) ~\cite{vill02,kiyo08,onum13,yama17}, and electron paramagnetic resonance (EPR) spectroscopy~\cite{vill02,kana17}.

\begin{figure}[t]
	\includegraphics[width=0.85\columnwidth]{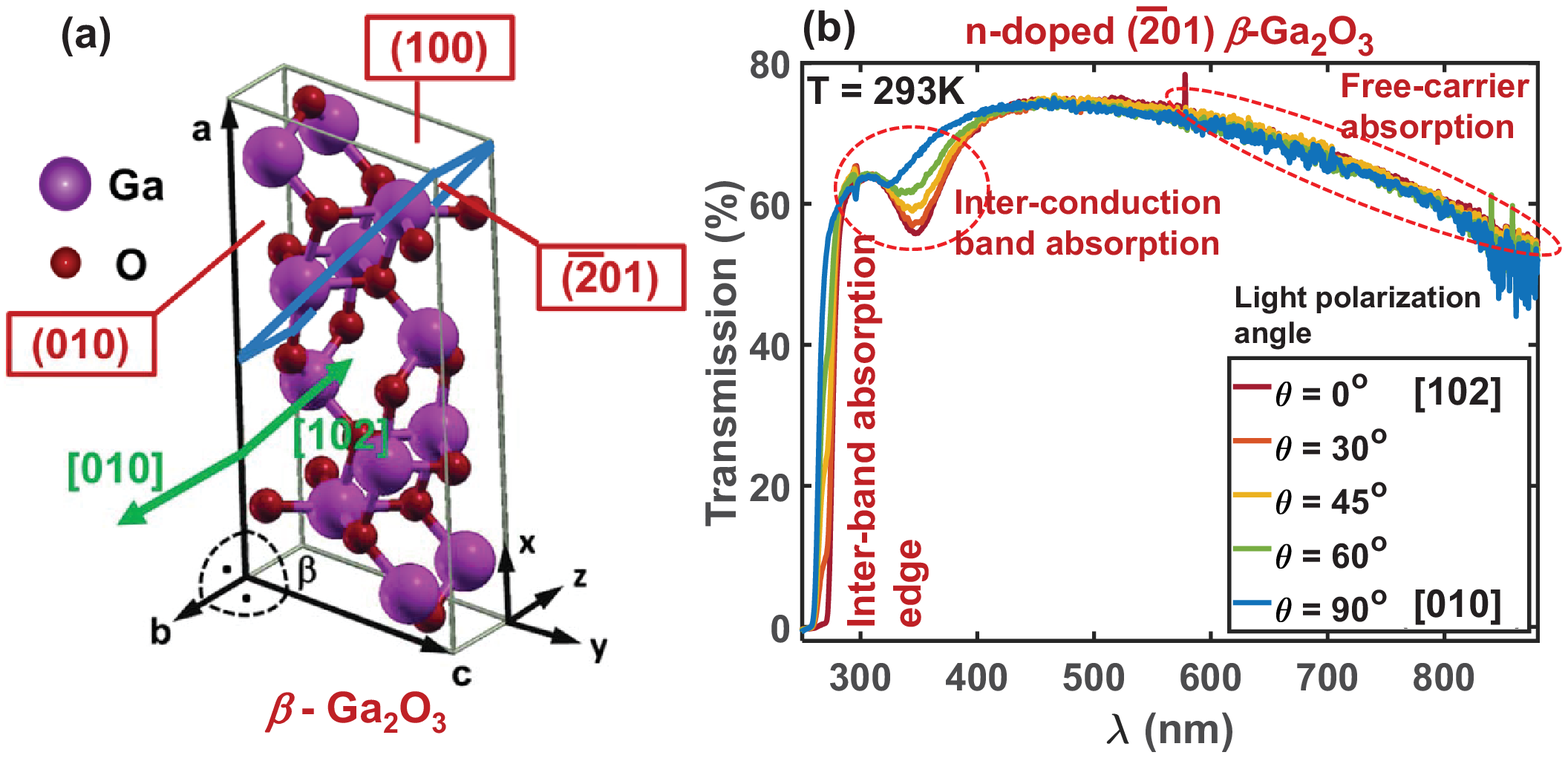}
	\caption{ (a) $\beta$-Ga$_2$O$_3$ unit cell depicting different crystal planes~\cite{Mock2017}. (b) Measured optical transmission spectrum of Sn-doped $(\bar{2}01)$ $\beta$-Ga$_2$O$_3$ for incident light polarized along the [102] and [010] crystal axes showing different light absorption mechanisms. Free-carrier absorption at near-IR wavelengths is used to probe photoexcited carrier dynamics in this work.}
        \label{fig:fig1}
\end{figure}

In this work, we report results from ultrafast optical pump-probe spectroscopy on bulk $\beta$-Ga$_2$O$_3$ crystals. We employ a two-color pump-probe scheme in which electrons are photoexcited from the valence band (VB) into the conduction band (CB) using a two-photon pump pulse, and the subsequent dynamics of the carriers are probed using a near-IR probe pulse whose transmission through the sample is affected by free-carrier IR absorption. Our results show that photoexcited electrons decay on nanosecond time scales. The measured transients are found to depend on the probe polarization with respect to the crystal [010] and [102]) crystal axes and suggest an additional probe absorption mechanism by photoexcited defect states. A carrier capture model based on rate equations is developed to extract the rate constants for electron and hole capture by defects. Whereas the rate constant for electron capture by defects is found to be temperature-independent, the temperature dependence of the hole capture rate constant can be fitted with a Mott-Seitz~\cite{alka14} expression and points to a lattice deformation/relaxation process accompanying hole capture. 

$\beta$-Ga$_2$O$_3$ bulk samples used in this work, obtained from the Tamura Corporation, were grown by edge-defined film-fed growth (EFG) technique~\cite{aida08,shim13,akit16}. We present results for a $300$ $\mu$m thick, Sn-doped, $(\bar{2}01)$ crystal with electron density $n \approx 10^{19}$ cm$^{-3}$ (see Supplementary Materials Section for similar results obtained for a $(100)$ $\beta$-Ga$_2$O$_3$ sample). The complex dielectric constant of the sample in the near-IR wavelength range is assumed to have the standard Drude form,
\begin{equation}
\label{eq:one}
\epsilon (\omega) = \epsilon(\infty) - i\frac{n e^2 \tau / m_e}{\omega (1+i \omega \tau)}
\end{equation}
Here, the high-frequency permittivity $\epsilon(\infty)$ is $3.85\epsilon_o$~\cite{indra11} and the CB electron effective mass $m_e$ is $0.27m_o$~\cite{hart15}. Hole contribution to the permittivity is neglected due to the predicted large effective masses of the VBs in $\beta$-Ga$_2$O$_3$~\cite{hai06,Frau2017}. Polarization-dependent optical transmission spectra for the n-doped sample are shown in Fig.~\ref{fig:fig1}(b) which shows the spectra obtained for light polarized between the [010] and the [102] axes. Hall measurements on the same bulk crystal showed an electron mobilty of $60$ cm$^2$/V-s. A high-frequency electron mobility of $155$ cm$^2$/V-s and a Drude scattering time $\tau$ of 22 fs were obtained by fitting (\ref{eq:one}) to the measured free-carrier absorption in the IR transmission data~\cite{SSM1967}.

\begin{figure}[t]
	\includegraphics[width=0.95\columnwidth]{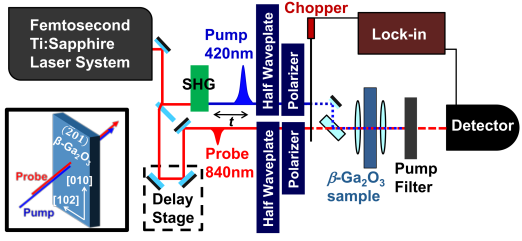}
	\caption{\label{fig:experiment} A schematic the two-color optical pump-probe setup. The inset shows the pump and the probe beam incident on the $(\bar{2}01)$ $\beta$-Ga$_2$O$_3$ sample.}
\end{figure}

Our two-color optical pump-probe scheme is depicted in Fig.~\ref{fig:experiment}(a). 840 nm center-wavelength, 65 fs pulse-width, optical pulses from a 83 MHz Ti:Sapphire laser were frequency doubled to generate 420 nm pump pulses that were used to photoexcite electrons from the VB of $\beta$-Ga$_2$O$_3$ to the CB by a two-photon absorption process. Transient absorption by photoexcited free-carriers of time-delayed 840 nm pulses probed the resulting carrier dynamics in the sample. Transmission of a near-IR probe pulse through the photoexcited sample is expected to be sensitive to free-carrier (intraband) absorption and not to band edge optical nonlinearities or carrier heating. Maximum pump-probe delay, limited by the mechanical delay stage, was 1.2 ns. Pump and probe beams were collinearly aligned and linearly polarized at different angles with respect to the crystal axes. Pump pulse fluences were in the 1-5 mJ/cm$^{2}$ range. Probe differential transmission $\Delta T/T$ as small as $10^{-6}$ can be measured in our scheme using lock-in detection. 

\begin{figure}
	\includegraphics[width=0.85\columnwidth]{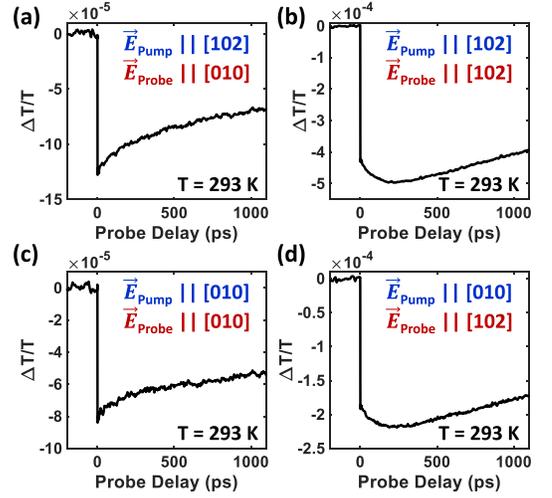}
	\caption{\label{fig:summary} 
	  Differential probe transmission $\Delta T/T$ is plotted as a function of the probe delay from the pump for different combinations of the pump and probe polarizations with respect to the [010] and [102] crystal axes of the $(\bar{2}01)$ $\beta$-Ga$_2$O$_3$ crystal. The polarization combinations are as indicated in the figure. Pump pulse fluence is $\sim$5 mJ/cm$^{2}$.} 
\end{figure}

The major results of the measurements are summarized in Fig.~\ref{fig:summary} which plots the measured $\Delta T/T$ as a function of the probe delay from the pump. The pump and probe pulses were linearly polarized either along the sample's [010] or [102] axes, leading to four polarization combinations. Rotating the pump polarization between these two axes, for constant probe polarization, had no qualitative effect on the shape of the measured transients, but rotating the \textit{probe} polarization had a significant effect. We explain this data below. 

\begin{figure}
	\includegraphics[width=0.95\columnwidth]{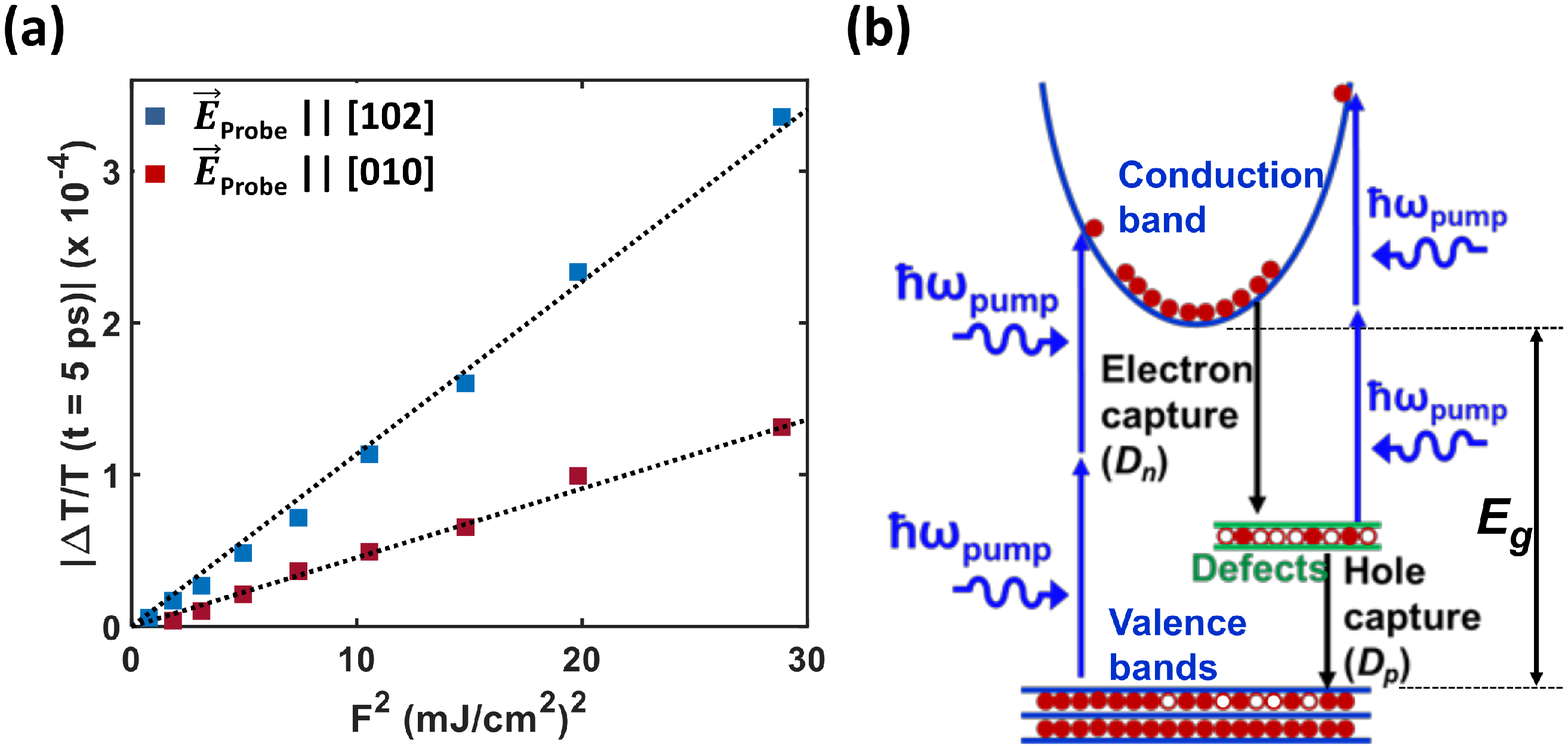}
	\caption{\label{fig:fluence} (a) Quadratic scaling of the measured $|\Delta T/T|$ signal (recorded $\sim$5 ps after photoexcitation) with the pump pulse fluence $F$ is shown, confirming photoexcitation via a two-photon process. Dotted lines are guides for the eye. The two curves are for the probe polarized along the [010] and [102] crystal axes. As discussed in the text, the [102] polarized probe also experiences absorption by defects depopulated by the pump. (b) Vertical arrows depict electronic transitions associated with photoexcitation from a two-photon absorption process and defect-assisted carrier recombination.}
\end{figure}

\begin{table}
	\caption{\label{tab:params}Extracted values of model parameters}
	\begin{ruledtabular}
		\bgroup
		\def\arraystretch{1.5}
		\begin{tabular}{ccc}
			Parameter&Value&Unit\\
			\hline
			$D_{n}$ & $\left(6.3\pm0.5\right)\,\times\,10^{-10}$ & cm\textsuperscript{3}s\textsuperscript{-1}\\
			$D_{p}$ & $\left(1.15\pm0.2\right)\,\times\,10^{-6}$ & cm\textsuperscript{3}s\textsuperscript{-1}\\
			$n_d$ & $\left(8\pm1\right)\,\times\,10^{14}$ & cm\textsuperscript{-3}\\
			$\sigma_{d}$ & $\left(6.5\pm1\right)\,\times\,10^{-17}$ & cm\textsuperscript{2}\\
			$F_{sat}$ & 6.62 & mJ/cm\textsuperscript{2}\\
		\end{tabular}
		\egroup
	\end{ruledtabular}
\end{table}

As in most other semiconductors, we expect the photoexcited electrons to thermalize and then relax to the CB bottom within a few ps~\cite{johan17}. Since the pump photon energy ($\sim$3 eV) is below the optical bandgap ($E_g$$\sim$4.7 eV), carrier photoexcitation is expected to be a two-photon process. The measured $|\Delta T/T|$ signal is expected to be from free-carrier absorption by the photoexcited electrons and is, therefore, proportional to the photoexcited electron density $\Delta n$. Results plotted in Fig.~\ref{fig:fluence}(a) show that $|\Delta T/T|$, recorded a few ps after photoexcitation, scales quadratically with pump pulse fluence $F$ as expected in the case of two-photon photoexcitation. 

Photoexcited electrons modify the permittivity and the complex refractive index $n^{2}_{r}(\omega) = \epsilon(\omega)/\epsilon_o$ (see (\ref{eq:one})) of the material, leading to higher transient absorption of the time-delayed probe pulse. Thus, $\Delta T/T$ is expected to be \textit{negative}, in agreement with our experimental results. As photoexcited electrons recombine and/or are captured by defects, $\Delta T/T$ recovers. Full recovery of the transient absorption is expected to last longer than $\sim$10 ns and exhibit multiple time constants, as ultrafast PL measurements have shown recently~\cite{yama17}. We focus on dynamics occurring within $\sim$1 ns. In this regime, $\Delta T/T$ increases monotonically from its value immediately after photoexcitation when the probe is polarized along the [010] axis. But when the probe is polarized along the [102] axis, $\Delta T/T$ decreases for $\sim$200 ps to a value less than that immediately after photoexcitation, and then increases in a manner similar to when the probe was polarized along the [010]. This initial decrease of $\Delta T/T$ in $\sim$200 ps when the probe is polarized along the [102] axis cannot be explained by free-carrier absorption, and suggests that an additional absorption mechanism is in play in the experiments. We present a defect-assisted carrier recombination model below that explains all features of the experimental data.

\begin{figure}
	\includegraphics[width=1.0\columnwidth]{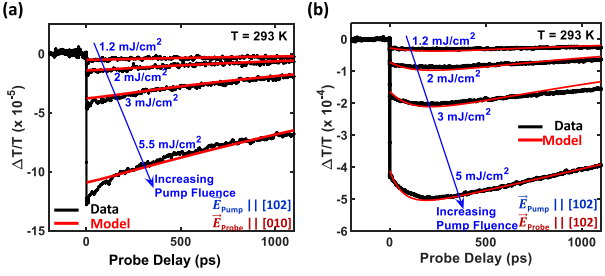}
	\caption{\label{fig:dataset} Results from the model are compared to the measured $\Delta T/T$ data for probe polarization along the (a) [010], and (b) [102] axis. The fitting parameters of the model are shown in the Table \ref{tab:params}.}
\end{figure}

Fig.~\ref{fig:fluence}(b) depicts our model for carrier photoexcitation by a two-photon absorption process and defect-assisted recombination. Defect to CB photoexcitation is also depicted as a two-photon absorption process and will be justified below. Following the Shockley-Read-Hall (SRH) model~\cite{SRH1952}, one can write the following rate equations for electrons and holes after photoexcitation,
\begin{eqnarray}
\label{eq:six}
\frac{dn}{dt} &=& -D_{n} \, n \, n_d \, (1-f_d) \\
\label{eq:seven}
n_d \frac{df_d}{dt} &=& D_{n} \, n \, n_d \, (1-f_d) - D_p \, p \, n_d \, f_d \\
\label{eq:eight}
\frac{dp}{dt} &=& -D_{p} \, p \, n_d \, f_d
\end{eqnarray}
Here, $n$ ($p$) is the electron (hole) density in the CB (VB), $n_{d}$ ($f_{d}$) is the density (occupation) of midgap defect states, and $D_{n}$ ($D_{p}$) is the rate constant for electron (hole) defect capture. Emission of carriers from defects is ignored for simplicity (assuming that midgap defect states are well separated in energy from band edges and emission times are much longer than the time scales probed in our experiments). Once the carrier densities have been computed, $\Delta T/T$ is obtained using $\Delta T/T \approx e^{-2\omega{\rm Im}\{\Delta n_{r}(\omega)\} L_{eff}/c} - 1$, where $\Delta n_{r}(\omega)$ is the refractive index change due to photoexcited carriers as given by (\ref{eq:one}), $L_{eff}\approx 160 \pm 15$ $\mu$m is the length over which the pump and probe beams interact in our setup, and $c$ is the speed of light. Using the previously measured values of the two-photon absorption coefficients of $\beta$-Ga$_2$O$_3$~\cite{Chen2018}, the change in electron/hole density after photoexcitation is $\Delta n = \Delta p = 2.5 \times 10^{14}F^{2}$ (or $1.4 \times 10^{14}F^{2}$) cm$^{-3}$/(mJ/cm$^2$)$^{2}$ for pump polarization along [102] (or [010]) axis. Defect occupation immediately after photoexcitation is assumed to have the form $1/(1+ F^2/F_{sat}^2)$, where $F_{sat}$ is used as a fitting parameter. Calculated and measured values of $\Delta T/T$ for probe polarization along the [010] axis for different pump pulse fluences agree well, as shown in Fig.\ref{fig:dataset}(a). Values of the fitting parameters, $D_{n}$, $D_{p}$, $n_{d}$, and $F_{sat}$, are given in Table \ref{tab:params}. The defect density value $n_{d}$ extracted from our measurements (see Table ~\ref{tab:params}) agrees well with densities (of the order of $n_{d}$ $\sim$ 10$^{15}$ cm$^{-3}$) obtained via other experimental techniques, such as DLTS and DLOS~\cite{Zhang2016}, for $\beta$-Ga$_2$O$_3$. The large difference in the values of $D_{n}$ and $D_{p}$ indicates that, for equal electron and hole densities, holes are captured by defects \textit{much faster} than electrons. The hole capture rate constant $D_{p}$ changes with temperature, as we show below.

Measured results for probe polarization along the [102] axis suggest an additional mechanism (besides free-carrier absorption) which increases probe absorption for $\sim$200 ps after photoexcitation. We propose that a polarization-selective VB-to-defect optical transition is the responsible mechanism. Immediately after photoexcitation, the defect levels are partially emptied and holes are also generated in the VB. Empty defect states allow optical transitions by the probe from the VB to the defect states. As time progresses, capture of the VB holes by defect states increases probe absorption. We model probe absorption by these VB-to-defect transitions with an additional absorption coefficient $\alpha_d = \sigma_d n_d (1-f_d)$ (units: 1/cm). The optical absorption cross section $\sigma_d$ is used as a fitting parameter. Additional absorption experienced by the probe when it is polarized along the [102] axis is also seen to scale quadratically with the pump fluence $F$, as shown in Fig.~\ref{fig:fluence}(a), which suggests that defect to CB photoexcitation is via two-photon absorption, as claimed earlier in the paper. 
Calculated and measured values of $\Delta T/T$ for probe polarization along the [102] axis are shown in Fig.\ref{fig:dataset}(b) for different pump pulse fluences and the agreement is found to be good for the same parameter values as the ones used in the case when the probe was polarized along the [010] axis. The extracted value of $\sigma_{d}$ (Table ~\ref{tab:params}) has the same order of magnitude as that of defects and impurities in more common group IV and group III-V semiconductors~\cite{Defects2013}. It is possible that the sample has more than one kind of midgap defect state but only one type of state participates in the VB to defect optical transitions. In this scenario, our model would provide only an average description and the fraction of defect states that contribute to probe absorption would be captured in the definition of the optical absorption cross section $\sigma_d$.

\begin{figure}
	\includegraphics[width=0.75\columnwidth]{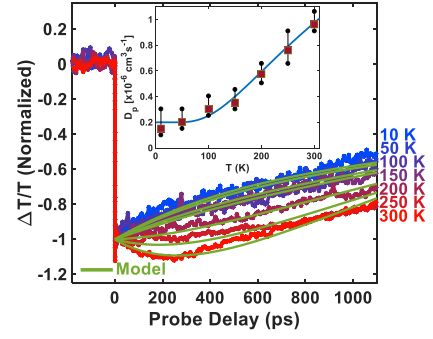}
	\caption{\label{fig:tempdep} Normalized $\Delta T/T$ transients, for probe polarized along the [102] axis, are plotted for different temperatures. Pump pulse fluence is 4 mJ/cm$^{2}$. A temperature-dependent hole capture rate constant $D_{p}$ explains the data extremely well. The value of $D_{n}$ (shown in Table ~\ref{tab:params}) is assumed to be temperature-independent. Inset: (Squares) Extracted values of $D_p$. (Solid curve) Fit to the data using a thermally-activated hole capture model~\cite{alka14}.}
\end{figure}

Temperature-dependence of carrier capture mechanisms can be insightful in determining their nature~\cite{Landsberg1992}. While radiative and Auger capture processes are relatively temperature independent, multi-phonon processes are strongly temperature dependent~\cite{Landsberg1992}. Measured $\Delta T/T$ transients showed no significant temperature dependence in the 10-300 K range when the probe was polarized along the [010] axis. However, the results were strongly temperature dependent when the probe was polarized along the [102] axis, as shown in Fig.~\ref{fig:tempdep}. With decreasing temperature, the initial decrease of $\Delta T/T$, within the first $\sim$200 ps disappears. A temperature-dependent $D_{p}$ explains this data very well. Fig.~\ref{fig:tempdep} shows the extracted values of $D_{p}$ as a function of the temperature. A Mott-Seitz expression of the form~\cite{alka14}, $D_p = C_0 + C_1 \exp (-\Delta E_b/k_B T)$, can be used to fit the observed temperature dependence for $C_0 = 2 \times 10^{-7}$ cm$^3$/s, $C_1 = 3 \pm 1 \times 10^{-6}$ cm$^3$/s, and $\Delta E_b = 35 \pm 5$ meV. The Mott-Seitz expression is commonly used to model thermally activated rates in multi-phonon processes~\cite{alka14,nonrad_book}.

Most common defect species in $\beta$-Ga$_2$O$_3$ are oxygen ($V_{O}$) and gallium ($V_{Ga}$) vacancies~\cite{Frau2017,Dong2017,Varley2010,Martin2013,Peel2016}. $V_{O}$ are predicted to have large overlaps with Ga:4s and O:2p orbitals of the CB and VB Bloch states, respectively~\cite{Dong2017}, making them favorable for fast and efficient carrier capture via radiative and Auger processes. Predicted energies for direct optical transitions between neutral $V_{O}$ and the CB are larger than our pump photon energy of 3.0 eV~\cite{Frau2017}, but such transitions would be allowed by the two-photon absorption process consistent with our observations. Absorption energies and polarization selection rules corresponding to optical transitions from the VB to $V_{O^+}$ have not been reported in the literature. However, emission energies for optical transitions from $V_{O}$ to the VB have been reported to be in the 1.4-2.25 eV range~\cite{Varley2010}, which is close to our probe photon energy. Finally, $V_{Ga}$ are known to have an O:2p character~\cite{kana17}, and optical transitions into them are expected to be allowed from deep within the VB where the orbitals have a small Ga:4s character~\cite{Dong2017}. In addition, the O:2p character of $V_{Ga}$ would make it efficient at hole capture, in agreement with our observations. Orientations of the O:2p orbitals either associated with the defect state or with the VB states could account for the polarization-selective probe absorption seen in our experiments. While the work presented here cannot completely distinguish between different intrinsic and extrinsic defects that can contribute to carrier capture, the sensitivity of the technique opens up pathways for more controlled future experiments.

In summary, we presented optical pump-probe spectroscopy results on defect-assisted recombination of photoexcited carriers in $\beta$-Ga$_2$O$_3$ and proposed a model that explains the measured data for different pump fluences and temperatures using a consistent set of parameters. Our results indicate a temperature-independent electron capture rate constant and a temperature-dependent hole capture rate constant indicating a thermally-activated hole capture process. Oxygen and gallium vacancies, common intrinsic defect states in $\beta$-Ga$_2$O$_3$, seem to be good candidates for defect states assisted carrier recombination. These studies of ultrafast carriers dynamics should prove useful in the design of high-voltage and high-speed transistors and photodetectors.

\begin{acknowledgments}
The authors would like to acknowledge fruitful discussions with Hartwin Peelaers, and partial supports from NSF under grant DMR-1120296, AFOSR under grant FA9550-17-1-0048, and NSF under grant DMR-1534303.
\end{acknowledgments}

\section{Supplementary Material}

In this Supplementary Material Section we present ultrafast pump-probe spectroscopy results for a bulk, 500 $\mu$m thick, $(100)$, unintentionally-doped, $\beta$-Ga$_2$O$_3$ crystal obtained from the Tamura Corporation. This crystal was also grown by edge-defined film-fed growth (EFG) technique~\cite{aida08,shim13,akit16}. The sample electron density and the Drude scattering time were obtained using optical transmission measurements in the near-IR to Mid-IR range, as described in the main text. The electron density $n$ and the Drude scattering time $\tau$ were estimated to be $6 \times 10^{17}$ 1/cm$^3$ and 22 fs, respectively. Given that at optical frequencies the Drude expression for the permittivity (Eq.(1) of the main text) is sensitive to only the ratio $n/\tau$, the electron density estimate is expected to be accurate to within a factor of 2 or 3. Hall measurements could not give more accurate results given the small electron densities.   

\begin{table}
	\caption{\label{tab:params_sup} (Supplementary) Extracted values of the model parameters for $(100)$  $\beta$-Ga$_2$O$_3$.}
	\begin{ruledtabular}
		\bgroup
		\def\arraystretch{1.5}
		\begin{tabular}{ccc}
			Parameter&Value&Unit\\
			\hline
			$D_{n}$ & $\left(2.3\pm0.5\right)\,\times\,10^{-9}$ & cm\textsuperscript{3}s\textsuperscript{-1}\\
			$D_{p}$ & $\left(2.1\pm0.12\right)\,\times\,10^{-6}$ & cm\textsuperscript{3}s\textsuperscript{-1}\\
			$n_d$ & $\left(2.8\pm0.4\right)\,\times\,10^{15}$ & cm\textsuperscript{-3}\\
			$\sigma_{d}$ & $\left(1.04\pm0.2\right)\,\times\,10^{-17}$ & cm\textsuperscript{2}\\
			$F_{sat}$ & 6.62 & mJ/cm\textsuperscript{2}\\
		\end{tabular}
		\egroup
	\end{ruledtabular} 
\end{table}

\begin{figure}
	\includegraphics[width=0.65\columnwidth]{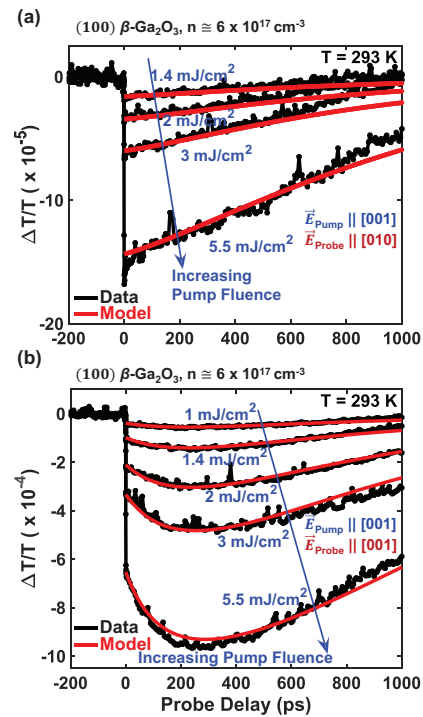}
	\caption{\label{fig:dataset_sup} (Supplementary) Results from the model are compared to the measured $\Delta T/T$ data for a $(100)$ $\beta$-Ga$_2$O$_3$ sample. Probe polarization is along the (a) [010], and (b) [001] axis. The extracted values of the model parameters are given in Table \ref{tab:params_sup}. Good agreement between data and the model is obtained with parameter values that are reasonably close the ones reported in the main text for the $(\bar{2}01)$ $\beta$-Ga$_2$O$_3$ sample. } 
\end{figure}

Fig.~\ref{fig:dataset_sup} summarizes the results of pump-probe measurements performed on the $(100)$ $\beta$-Ga$_2$O$_3$ sample. Using the defect-assisted recombination model introduced in the main text, the extracted values of  $D_{n},D_{p},n_{d},\sigma_{d}$, and $F_{sat}$ are summarized in Table \ref{tab:params_sup}. The values obtained for $D_{n}$, $D_{p}$ , $n_{d}$, and $F_{sat}$ are within a factor of 2 or 3 of the corresponding values reported in the main text for the $(\bar{2}01)$ $\beta$-Ga$_2$O$_3$ sample. The small discrepancy is attributed to the accuracy with which we were able to estimate the equilibrium electron density of the $(100)$ sample. The extracted value of the defect optical absorption cross-section $\sigma_{d}$ is almost 5 times larger for the $(\bar{2}01)$ sample compared to the $(100)$ sample. Given that the probe polarization was along the [001] and [102] in the $(100)$ and  $(\bar{2}01)$ samples, respectively, the strengths of the optical matrix elements are expected to be different. In addition, any error in the determination of the defect density $n_{d}$ could also contribute an error to the extracted value of $\sigma_{d}$ since it is the product $\sigma_{d} \, n_{d}$ that determines optical absorption. Interestingly, the extracted values of the product $\sigma_{d} \, n_{d}$ are within a factor of 2 in the two samples.

\bibliography{Ga2O3_library_v5}

\end{document}